\documentstyle[epsfig,aps,pre,floats,tighten]{revtex}
\begin{document}
\draft
\title{I. Territory covered by $N$ random walkers on deterministic fractals. \\ The Sierpinski gasket}
\author{L. Acedo and S. B. Yuste}
\address{Departamento de F\'{\i}sica, Universidad  de  Extremadura,\\
E-06071 Badajoz, Spain}
\date{\today}
\maketitle
\begin{abstract}
We address the problem of evaluating the number $S_N(t)$ of
distinct sites visited up to time $t$ by $N$ noninteracting random
walkers all initially placed on one site of a deterministic fractal
lattice. For a wide class of fractals, of which the Sierpinski gasket is a typical example, we propose that, after the short-time
compact regime and for large $N$, $S_N(t) \approx \widehat{S}_N(t)
(1-\Delta)$, where $\widehat{S}_N(t)$ is the number of sites inside a
hypersphere of radius $R [\ln (N)/c]^{1/ u }$,  
$R$ is the root-mean-square displacement of a single random walker, 
and $u$ and $c$ determine how fast $1-\Gamma_t({\bf r})$ (the probability that site ${\bf r}$ has been visited by a single random walker by time $t$) decays for large values of $r/R$: $1-\Gamma_t({\bf r})\sim \exp[-c(r/R)^u]$. For the deterministic fractals considered in this paper,  $ u =d_w/(d_w-1)$, $d_w$ being the random walk dimension. 
The corrective term $\Delta$ is expressed as a series in  $\ln^{-n}(N) \ln^m \ln (N)$ (with $n\geq 1$ and $0\leq m\leq n$), which is given explicitly up to $n=2$. Numerical simulations on the Sierpinski gasket show reasonable agreement with the analytical expressions. 
The corrective term $\Delta$ contributes substantially to the final value of $S_N(t)$ even for relatively large values of $N$.
\end{abstract}

\section{INTRODUCTION}
\label{sect_1}
Random walk theory is a branch of statistical physics with many applications
\cite{HughesWeiss,SHDBA}. Problems related to a single random walker have
traditionally been the subject of thorough study, but their generalizations to $N > 1$ random walkers have attracted much less attention, although there are  some, generally very recent, exceptions  
\cite{Larral1,Larral2,Havlin,fewN,RC,PREeucl,DK}. 
These  multiparticle diffusion problems are characterized by the  impossibility of being analyzed in terms of the single random walker theory, i.e., they can not be solved through simple averaging over the properties of a single random walker, even in the noninteracting case.
The recent development of experimental techniques allowing the observation of events caused by single particles of an ensemble \cite{SingMol} should give additional encouragement to the study of these of multiparticle diffusion problems. 

The problem that is the subject of this paper, namely, the evaluation of the average number $S_N(t)$ of distinct sites visited (or territory covered) by $N$ random walkers up to time $t$, all moving from the same starting site, is a clear example of a diffusion problem that can not be solved, or even approximated, from the solution for $N=1$, $S_1(t)$. Even for independent random walkers, the overlap of the regions explored by different walkers prohibits a decomposition of $S_N(t)$ into single-particle contributions.
The origin of the problem of evaluating $S_N(t)$ is usually traced back to the Proceedings of
the Second Berkeley Symposium on Mathematical Statistics and Probability
where the case $N=1$ was posed by Dvoretzky and Erd\"os \cite{DE}. Since then, the quantity $S_1(t)$ has been studied in detail and is discussed in general references \cite{HughesWeiss,Montroll}. 
For fractal substrates (in particular, for the two-dimensional Sierpinski gasket and the  two-dimensional percolation cluster at criticality) this problem was first studied by Rammal {\em et al.} \cite{SHDBA,Rammal}. More recently, Larralde {\em et al.} \cite{Larral1,Larral2} and Havlin {\em et al.} \cite{Havlin} studied the problem of evaluating $S_N(t)$ when $N \gg 1$ noninteracting random walkers diffuse in Euclidean and fractal media, respectively.  
For fractal lattices with spectral dimension $d_s=2 d_f/d_w < 2$,
 it was argued  \cite{Havlin} that
$S_N(t)\sim t^{d_\ell}$ for $t<t\ll t_{\times} \sim \ln N$ and 
$S_N(t)\sim t^{d_s/2} \ln^{d_f/u}\left( N \right)$ for 
$ t_{\times} \ll t $,
where $d_{\ell}=d_f/d_{\text{min}}$ is the chemical dimension (or topological distance dimension), 
 $d_f$ is the fractal dimension, $d_{\text{min}}$ is the fractal dimension of the shortest path on the fractal,  
$u=d_w/(d_w-1)$ and $d_w$ is the diffusion exponent (or fractal dimension of the random walk) \cite{SHDBA,BH}.
However,  Dr\"ager and Klafter \cite{DK} using scaling arguments have recently proposed that 
\begin{equation}
\label{SNtHav}
S_N(t) \sim 
t^{d_s/2} (\ln N)^{d_\ell/v}
\end{equation}
for $t_{\times} \ll t$, 
where $v=d_w^\ell/(d_w^\ell-1)$ and $d_w^\ell=d_w/d_{\text{min}}$ is the chemical-diffusion exponent \cite{SHDBA,BH}. 
Of course, the two predictions agree for the media considered, such as Sierpinski gaskets, in the present paper for which $d_{\text{min}}=1$. 
Therefore, in what follows, we will write $d_f$, $u$ and $d_w$ instead of $d_\ell$, $v$  and $d_w^\ell$, respectively. 
Fractals with $d_{\text{min}}\neq 1$ will be discussed in the following paper \cite{paperII}.

 As stated above, two time regimes are observed in $S_N(t)$: an extremely short-time regime or regime I and a long-time regime or regime II separated by the crossover time $t_{\times} \sim \ln N$. 
A further long-time regime, or regime III, is observed in Euclidean lattices when the movement of the independent walkers are very far from each other so that their trails (almost) never overlap and $S_N(t) \sim N S_1(t)$ \cite{Larral1,Larral2,RC,PREeucl}. 
In the one-dimensional lattice and fractal lattices with $d_s \le 2$, the trails of the random walkers partially overlap at all times and regime III is never reached. 
Such is the case in this paper where we are concerned only with fractals in which $d_s < 2$. The transition from regime I to regime II is easy to understand. In regime I, we have so many particles at every site that all nearest neighbors of the already visited sites are necessarily reached by some walker at the next time step. The minimum path length between two sites on a fractal is the chemical distance, $\ell$, and it is clear that after a time $t$ the visited zone is an hypersphere of chemical radius $\ell=t$
and, consequently, $S_N(t) \sim t^{d_{\ell}}$ according to the definition
of the chemical exponent \cite{SHDBA,BH}. This behavior lasts until the average
number of particles per site is of the order of unity, then overlapping
is only partial and a new regime is established. If $z$ is the coordination
number of the lattice (or the average coordination number in the case of stochastic fractals), the number of random walkers at every site decreases as $N/z^t$ in the very short-time regime because the particles are distributed between $z$ sites at every step. 
The breaking of the overlapping regime thus takes place when $N/z^t \sim 1$ or, equivalently, $t_{\times} \sim \ln N$. 

Regime II is far more interesting and difficult to analyze than regime I due to the nontrivial interplay of the walkers in their exploration of the lattice, which leads to a more complex dependence of $S_N(t)$ on time $t$ and number of particles $N$. In some recent work \cite{RC,PREeucl}, we have shown that for independent random walks on Euclidean lattices there exist important asymptotic corrections to the main term  that can not be ignored even for very large number of particles as these corrections decay only logarithmically. We will see that this also holds for the two-dimensional Sierpinski gasket.
 
It should be noticed that, except for the Sierpinski gasket in two dimensions when $N=1$ \cite{Rammal}, there has never been any discussion about $S_N(t)$ focused on deterministic fractals, whether theoretically or numerically.
Certainly, a dependence on $t$ and $N$ of the main asymptotic term of $S_N(t)$ for large $N$  has been proposed [see Eq.\ (\ref{SNtHav})], but nothing is known about the value of its amplitude or prefactor and on the relevance (if any) of the other (corrective) asymptotic terms.
In this paper we present a procedure for obtaining, for a certain class of fractals, the complete asymptotic series expansion of $S_N(t)$ when $N\gg 1$. The procedure gives the main asymptotic term in full, and determines the functional form of the corrective terms, which  we calculate explicitly up to second order. 
The fractals that we consider in this paper have to satisfy two conditions.  First, the number of sites (or volume) $V(r)$ of the fractal inside a hypersphere of radius $r$ should be given by
\begin{equation}
\label{Vr}
V(r)=V_0 r^{d_f}
\end{equation}
where $V_0$ is a constant characteristic of the fractal substrate; and, second, the probability  $\Gamma_t({\bf r})$ that a site ${\bf r}$ has not been visited by a single random walker by time $t$ should decay for  $\xi= |{\bf r}|/R\equiv r/R \gg 1$ as 
\begin{equation}
\Gamma_t({\bf r})\approx 1-A \xi^{-\mu u} \exp(-c\xi^u)
\left(1+ h_1 \xi^{-u}+ \cdots\right),
\label{gasinSimple}
\end{equation}
where $R^2=2 D t^{2/d_w}$ is the mean-square displacement of a single random walker and $D$ is the diffusion constant.
It should be clear at this point that the above two conditions can only be  satisfied approximately: first, because $V_0$ is not strictly constant (it exhibits  log-periodic oscillations of small amplitude, see Sec.\ \ref{sect_3}); second, because  $\Gamma_t({\bf r})$ does not solely depend on the distance $r$ but also (in general) on the actual location ${\bf r}$ on the lattice; and, third, because $\Gamma_t({\bf r})$ is not continuos  (this fact can be clearly seen in figure 1 of Ref.\ \cite{Bidaux}) so that Eq.\ (\ref{gasinSimple}) can only be an approximation to the true distribution. The fluctuations in $S_N(t)$ associated to these effects are thus not included in our theoretical discussion. However, their importance can be gauged by resorting to simulation.  For the two-dimensional Sierpinski gasket, we found that these fluctuations are indeed relevant and that they can be explained to a large extent as a consequence of the log-periodic oscillations of $V_0$. 
Finally, there is another difficulty regarding the value of $\Gamma_t({\bf r})$: while its dominant term $\exp(-c\xi^u)$ is reasonably well established, the value (and even the form)  of its subdominant factors $A \xi^{-\mu u}$, $h_1 \xi^{-u}$, etc.\ is unknown (although an educated guess can be made; see Sec.\ \ref{sect_2}).  This means that we can be fairly sure of the value of the main term of $S_N(t)$ because, as we will show, it depends only on the dominant term of $\Gamma_t({\bf r})$. However, the true value of the corrective terms of $S_N(t)$ is more uncertain as they also depend on the subdominant factors of $\Gamma_t({\bf r})$. 
Nevertheless, we will see that reasonable choices of values for these subdominant factors lead to significant improvements in the estimate of $S_N(t)$.

The simulation results for the two-dimensional Sierpinski gasket will reveal the great importance of the corrective terms (for example, they account for some thirty per cent of the total value of $S_N(t)$ even for $N$ as large as $10^6$).  This is not strange because, as we will show in Sec.\ \ref{sect_2}, they decay only logarithmically with $N$.
 An important consequence that we will address in the following paper \cite{paperII} is that the corrective terms  must be taken into account in the analyses based on ``collapsing'' the numerical data \cite{Larral1,Havlin,DK} to determine the exponents in the main term of $S_N(t)$. 
 
The  paper is organized as follows. The asymptotic evaluation of $S_N(t)$ on fractal lattices is presented in Sec.\ \ref{sect_2}. The mathematical techniques involved in the calculation are very similar to those corresponding to the Euclidean case and we will only outline the main steps. Details may be found in Ref.\ \cite{PREeucl}. In Sec.\ \ref{sect_3}, we compare the asymptotic expansion of $S_N(t)$  with simulation results obtained on the Sierpinski gasket.
We end in Sec.\ \ref{sect_4} with some remarks on the quality of the asymptotic approximation.

\section{TERRITORY COVERED BY $N$ RANDOM WALKERS ON A DETERMINISTIC FRACTAL SUBSTRATE}
\label{sect_2}
\begin{figure}
\begin{center}
\parbox{0.4\textwidth}{
\epsfxsize=\hsize \epsfbox{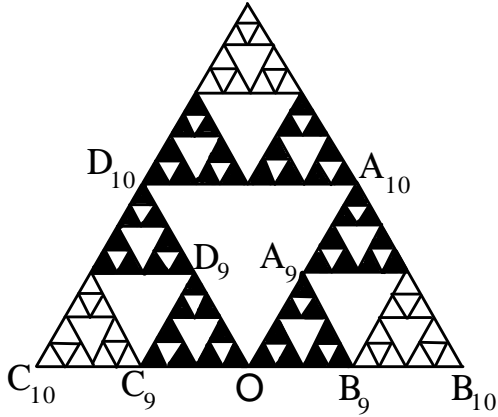}}
\caption{ The $11$-generation Sierpinski lattice used in the numerical simulations where the smallest triangles represent for $8$-generation lattices. 
The sites labelled $A_i$, $B_i$, $C_i$ and $D_i$ are nearest neighbors of O in the $i$-th times decimated lattice \protect\cite{yusteJPA,broeck}.
In case I all random walkers start from the origin O.
In case II the common origin is chosen randomly from among the sites in the shaded area. 
} 
\label{figSierpiRed}
\end{center}
\end{figure}

We will define the multiparticle survival probability $\Gamma_N(t,{\bf r})$ as the probability that site ${\bf r}$ has not been visited by time $t$ by any of the $N$ random walkers that start from the origin site ${\bf r}={\bf 0}$ at time $t=0$. 
From this definition, the following relationship between the average number of distinct sites visited, $S_N(t)$, and the survival probability can be established \cite{Larral1,Larral2}:
\begin{equation}
\label{SG}
S_N(t)=\sum_{\bf r}\, \left\{ 1 - \Gamma_N(t,{\bf r}) \right\}\; ,
\end{equation}
where the sum is over all the sites in the fractal lattice.
For independent random walkers, we have $\Gamma_N(t,{\bf r})=\left[
\Gamma_t ({\bf r}) \right]^N$, where $\Gamma_t ({\bf r})\equiv \Gamma_1(t,{\bf r})$ is the one-particle
survival probability. Since we are interested in the behavior of $S_N(t)$ after a large number of steps (thus beyond of the very-short-time regime I), we
replace $\Gamma_t({\bf r})$ and $S_N(t)$ by their continuum approximations:
\begin{eqnarray}
\label{SGC}
S_N(t)&=&\displaystyle\int_0^\infty\,\left[ 1 - \Gamma_t^N(r) \right] dV(r)\\
\noalign{\smallskip}
 &=&V_0 (2 D)^{d_f/2}\, t^{d_f/d_w}\, J_N(d_f)\; ,
\end{eqnarray}
where we have assumed that the fractal volume $V(r)$ (i.e., the number of lattice sites) of the hypersphere with radius $r$ is given by Eq.\ (\ref{Vr}) and that, essentially,  $\Gamma_t({\bf r})$  only depends on the distance $r$. In Eq.\ (\ref{SGC}), $J_N(d_f)$ is
\begin{equation}
\label{JNab}
J_N(d_f)= N \int_0^\infty \,\left[\Gamma_{t}(\xi)\right]^{N-1}
\frac{d \Gamma_t(\xi)}{d \xi} \xi^{d_f} d \xi  .
\end{equation}
In order to evaluate this integral for $N\gg 1$ it suffices to know
$\Gamma_t(r)$ for large $\xi$.  For Euclidean lattices  
\begin{equation}
\Gamma_t(r)\approx \Gamma(\xi)\equiv 1-A \xi^{-\mu u} \exp(-c\xi^ u )
\left(1+\sum_{n=1}^\infty h_n \xi^{-n u }\right)
\label{gasin}
\end{equation}
when $\xi\gg 1$ \cite{Larral2,RC,PREeucl}. 
In this paper we shall consider only those fractals, such as the Sierpinski gaskets, for which this equation approximately [see discussion below Eq.(\ref{gasinSimple})] holds. 
 Indeed, we can show that Eq.\ (\ref{gasin}) is reasonable for those lattices ($d$-dimensional Sierpinski fractals, Given-Mandelbrot curve,  one-dimensional lattice, hierarchical diamond lattice,\ldots) where the renormalization procedure implemented in Refs. \cite{yusteJPA,Porra,broeck} can be set up.
The argument is as follows.
Let ${\bf r}^{(i)}_n$, $n=1,\ldots,z$, with $r^{(i)}=|{\bf r}^{(i)}_n|$, be the position of the $z$ nearest neighbors of the site at $\bf{r}=\bf{0}$ in the fractal lattice decimated $i$ times (see Fig.\ \ref{figSierpiRed}), and let $h(t,r^{(i)})$ be the probability that, in the time interval [0,t], a single diffusing particle that starts at $\bf{r}=\bf{0}$ is absorbed by {\em any} of the traps located at its $z$ nearest neighbors placed at the sites ${\bf r}^{(i)}_n$.
For large values of $\xi\sim r^{(i)}/R$, i.e., for relatively short times, the event
``the random walker arrives for the first time at site ${\bf r}^{(i)}_n$" and the event
``the random walker arrives for the  first time at site ${\bf r}^{(i)}_m$" are (almost) independent
so that $h(t,r^{(i)})$ is (almost) the sum of the probability [given by $1-\Gamma_t(r^{(i)})$] of each of the $z$ individual events, i.e., 
$h(t,r^{(i)})\simeq z \left[1-\Gamma_t(r^{(i)})\right]$. 
Of course, the two events are not fully independent because the random walker could first arrive at site ${\bf r}^{(i)}_m$ after passing by the site ${\bf r}^{(i)}_n$.
However, this is very unlikely because ${\bf r}^{(i)}_m$ and ${\bf r}^{(i)}_n$ are separated by distances of order of $r^{(i)}$ so that the fraction of random walkers that, after arriving at ${\bf r}^{(i)}_n$, travel to ${\bf r}^{(i)}_m$ in the short time interval [0,t] is of the order $\exp(-\xi^ u )$, with $\xi\sim r^{(i)}/R\gg 1$, because  the propagator or Green's function is of this order for large values of $\xi$ \cite{yusteJPA,yusteacedoPRE}. Therefore, it is reasonable to assume that
$1-\Gamma_t(r^{(i)})=h(t,r^{(i)})/z \{1+{\cal O}(\exp(-\xi^ u )]\}$ for large $\xi$.
But $1-h(t,r^{(i)})$ has the form of the right-hand side of Eq.\ (\ref{gasin}) with $u=d_w/(d_w-1)$, at least for the fractals that we are considering \cite{yusteJPA,broeck}, so that Eq.\ (\ref{gasin}) for ${\bf r}={\bf r}^{(i)}$ follows. 
In this discussion we have assumed, in order for the renormalization analysis that leads to $\Gamma_t\left(r^{(i)}\right)$ to work,  that the traps were placed at those sites (such as $A_g$,$B_g$,\ldots in Fig.\ \ref{figSierpiRed}) that become the nearest neighbors of the starting site after several decimations. But, in the calculation of $S_N(t)$, the function $\Gamma_t(r)$ is required for every pair of origin and destination sites in the lattice.
At this point, we shall assume that the survival probability for ${\bf r}\neq{\bf r}^{(i)}$ and ${\bf r}={\bf r}^{(i)}$ are very similar, i.e., we assume that $\Gamma_t(r^{(i)})\simeq \Gamma_\tau(r) = \Gamma_\tau(\xi)$ when $r/\tau^{1/d_w}=r^{(i)}/t^{1/d_w}=(2D)^{1/2}\xi$ for large $\xi$,  with $\Gamma_t(\xi)$  given by Eq.\ (\ref{gasin}). 
 To the best of our knowledge this problem has not been studied and will require a specific and detailed simulation analysis that is not the object of the present work. Nevertheless, previous simulations on the Sierpinski gasket of other statistical quantities closely related to $\Gamma_t(r)$, such as the propagator (or Green's function) \cite{yusteacedoPRE}, enable us to affirm with confidence that the parameters $c$ and $u$ remain unchanged over the whole Sierpinski lattice whereas the subdominant ones, $A$, $\mu$, $h_1$,\ldots, do not. 
In the following, we use $c=0.981$ \cite{yusteJPA,yusteacedoPRE} and $u=d_w/(d_w-1)\simeq 1.756$ for the two-dimensional Sierpinski gasket  (values of $c$ for other fractals can be found in Ref.\ \cite{yusteJPA}) but we must bear in mind that the actual values of $A$, $\mu$ and $h_n$, $n=1,2,\ldots$ will likely differ from those obtained by renormalization techniques (namely, $A=0.61 $, $\mu=1/2$, $h_1=-0.56$) as these latter correspond to the special placing of the traps and origins. 
We must also point out that, as shown below, since the parameters $h_n$, $n=1,2,3\ldots$, only contribute to the second and higher order series terms of $S_N(t)$  and since the real value of even the first-order asymptotic term is uncertain, the values of these parameters will not be considered here.

The evaluation of $J_N(d_f)$ now proceeds in analogy with
the analysis for the Euclidean lattices \cite{RC,PREeucl} and we shall only quote the final result for $S_N(t)$ in the fractal case:
\begin{equation}
S_N(t) \approx \widehat{S}_N(t) (1-\Delta)
\label{SNt}
\end{equation}
with 
\begin{eqnarray}
\label{hatSNt}
\widehat{S}_N(t)&=&V_0 \left( 2 D \right)^{d_f/2} t^{d_s/2} \left(\frac{\ln N}{c}
\right)^{d_f/ u } \\
\noalign{\smallskip}
\label{Delta}
\Delta &=&\beta \displaystyle\sum_{n=1}^\infty\, \ln^{-n} 
N \,
\displaystyle\sum_{m=0}^n \, s_m^{(n)} \ln^m \ln N \;, 
\end{eqnarray}
and where, up to second order ($n=2$),
\begin{eqnarray}
\label{s10}
s_0^{(1)}&=&-\omega  \\
s_1^{(1)}&=&\mu   \\
s_0^{(2)}&=&
-(\beta-1) \left( \frac{\pi^2}{12}+\frac{\omega^2}{2} \right) -
(c h_1-\mu \omega) \\
s_1^{(2)}&=& -\mu^2 + (\beta-1)\mu \omega \\
s_2^{(2)}&=& -\frac{1}{2} (\beta-1) \mu^2
\label{stilde}
\end{eqnarray}
and $\omega=\gamma+\ln A+ \mu \ln c$, $\gamma \simeq 0.577215$ is the
Euler constant, and $\beta=d_f/ u =d_f (1-1/d_w)$.

\section{NUMERICAL RESULTS FOR THE SIERPINSKI GASKET}
\label{sect_3}
To check the reliability of the analysis presented in the
previous section we carried out simulations of the number of
distinct sites visited by $N$ random walkers on a two-dimensional Sierpinski lattice with $g=11$ generations. This means that if we take the length of any side of the smallest triangles (the zeroth decimated triangles) as the unit length,
then the length of the sides of the triangle that inscribes the lattice (the $g$th decimated triangle) is $2^g$. Two different cases are analyzed: (I) random walkers are initially placed upon the center of the base of the main triangle which inscribes the lattice (point O in Fig. \ref{figSierpiRed}), and 
(II) the common starting site is randomly selected.
Qualitatively and quantitatively, the results are different in case I and  case II. The structure of the lattice gives rise to oscillations superimposed on the general trend of $S_N(t)$ in case I. This structure is smeared out in case II by the double average over experiments and over starting sites, so that $S_N(t)$ is now a smooth function.

\begin{figure}
\begin{center}
\parbox{0.4\textwidth}{
\epsfxsize=\hsize \epsfbox{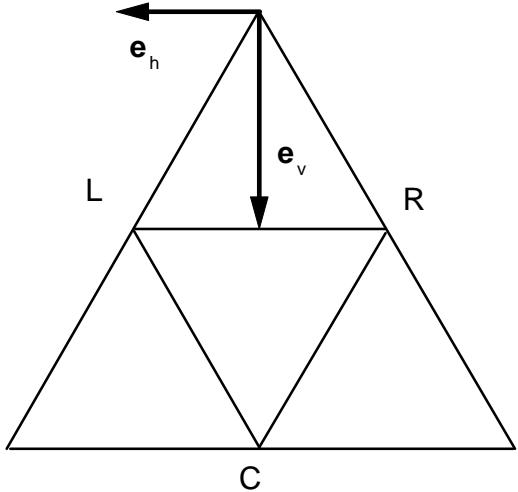}}
\caption{
Basis vectors used in the simulations to identify the lattice sites of
the two-dimensional Sierpinski lattice. The labels $R$, $L$ and $C$
correspond to the three different site types according to the relative
position of their neighbors.}
\label{figBasisVectors} 
\end{center}
\end{figure}

We shall use the top vertex of the main triangle as the origin of the lattice and the orthogonal basis $\{ {\bf e}_v,{\bf e}_h \}$ shown in Fig.\ \ref{figBasisVectors}, because it is convenient to divide the lattice into 
horizontal sets of sites with a fixed $v$ coordinate. The positions of the
random walkers are updated at every time step by a random selection of the
destination sites among the neighbors of the site occupied by the walker. In order to do that we have to take into account that the sites of the Sierpinski lattice may be classified into three types according to the relative positions of their neighbors: type $R$, $L$ and $C$ as shown in Fig.\ \ref{figBasisVectors}. 
Since the Sierpinski lattice has many holes on every scale, it is not efficient to identify the lattice sites and the random walkers positions by two coordinates $\{v,h\}$ because there are many coordinate pairs corresponding to no actual lattice site. 
Much memory can be saved if the sites are numbered from top to bottom and from left to right, so that the top vertex of the main triangle is site number $1$ and the right vertex is site number $3(3^g+1)/2$.

\subsection{Case I}

\begin{figure}
\begin{center}
\parbox{0.4\textwidth}{
\epsfxsize=\hsize \epsfbox{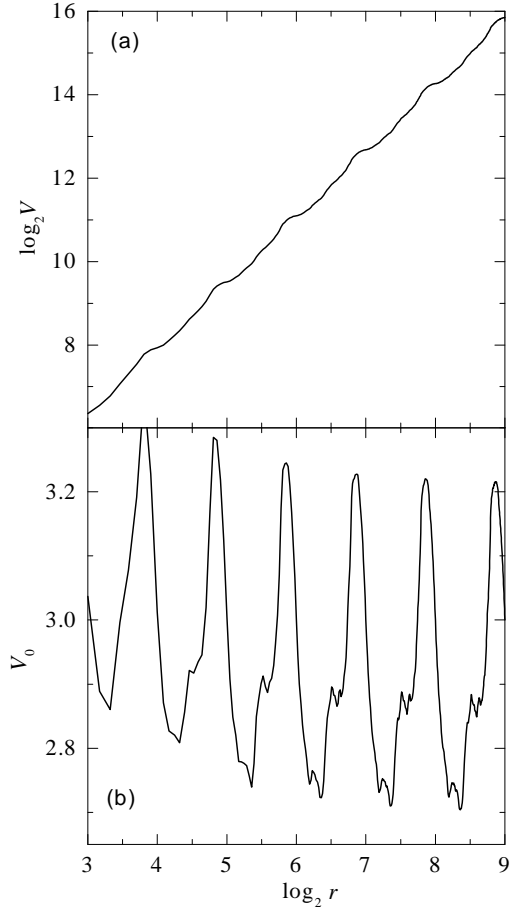}}
\caption{The fractal volume $V(r)$ of the Sierpinski lattice as a function of $r$ for case I.   (a) $\log_2 V(r)$ versus $\log_2(r)$; (b) $V_0=V(r)/r^{d_f}$ versus $\log_2(r)$. }
\label{figVr}
\end{center}
\end{figure}

First, we will discuss the simulation results for the territory covered
by $N$ random walkers placed initially at site O in Fig.\ \ref{figSierpiRed}. In
order to compare with the zeroth- and first-order asymptotic expression, cf. Eq.\  (\ref{SNt}), we must know the values of  $V_0$, $D$, $c$, $u$, $A$ and $\mu$. 
 In Fig.\ \ref{figVr}(a) we have plotted the
fractal volume of a circle of radius $r$, $V(r)$, centered upon the privileged site O. The observed log-periodic structure is a consequence of the empty and filled triangular areas that repeat periodically as $r$ is increased, but the general trend is well represented by a term of the form $V_0 r^{d_f}$ with $V_0\simeq 3.0\pm 0.1$.   In Fig. \ref{figVr}(b), in which the quotient $V(r)/r^{d_f}$ is plotted versus $\log_2r$, one clearly sees the log-periodic oscillations of $V_0$.
As our theory assumes a constant value for $V_0$, we take the average value over the last period (from maximum to maximum), $V_0=2.93$, as a reasonable criterion for comparison with the simulation results.  
In order to find the diffusion coefficient $D$ of a random walker starting at O, we performed $10^6$ simulations up to $t=400$.
The linear numerical fit between $t=50$ and $t=400$  gives $d_w \simeq 2.32$ (the exact value is $d_w=\ln 5/\ln 2 \simeq 2.322$) and $2 D \simeq 1.05$. 
Numerical fits using other time intervals (excluding short times, of course) lead to similar values, and we take $2 D \simeq 1.05\pm 0.02$ as a reliable estimate.
For the parameters $c$ and $u$ we take the values (see Sec.\ \ref{sect_2})  $0.981$ and $d_w/(d_w-1)=1.756$, respectively.
As discussed in Sec.\ \ref{sect_2}, the values of  $\mu$ and  $A$ are much less certain and we will use here two pairs of values: those obtained by renormalization, i.e.,  $\mu=1/2$ and $A=0.61$, and these same values increased by one, i.e., $\mu=3/2$ and $A=1.61$. Of course, this last pair  of parameters are arbitrary (other values could also be used) and are mainly given to show the relevance of the corrective terms.

\begin{figure}
\begin{center}
\parbox{0.4\textwidth}{
\epsfxsize=\hsize \epsfbox{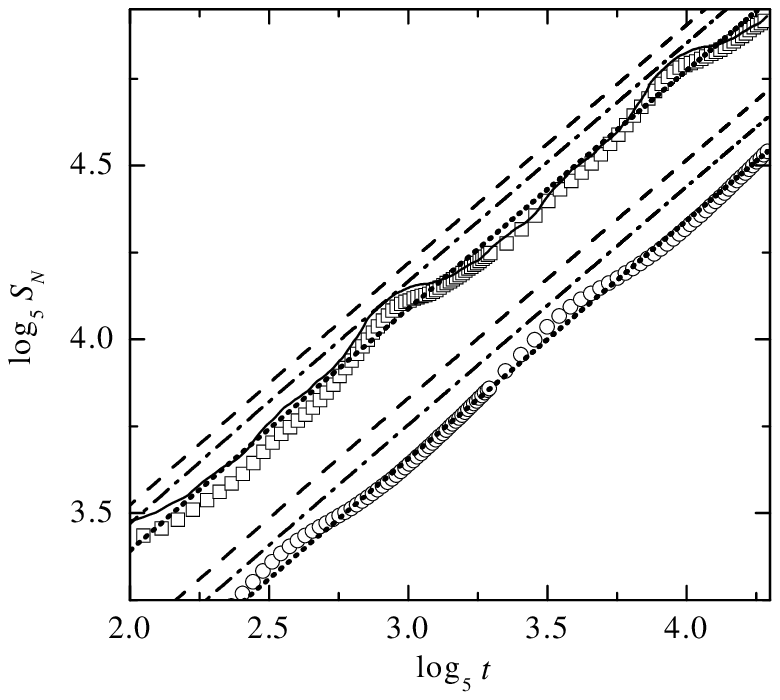}}
\caption{The number of distinct sites visited on the Sierpinski lattice until $t=1000$ by $N=10^3$  (circles) and $N=10^6$ (squares) random walkers. All random walkers start from the origin O shown in Fig.\ \protect\ref{figSierpiRed}.
The dashed line is the zeroth-order approximation, the dot-dashed line  corresponds to the first-order approximation with $\mu=1/2$ and $A=0.61$ and the  dotted line is  the first-order approximation with  $\mu=3/2$ and $A=1.61$. The solid line also corresponds to this last approximation but using for $V_0$ the numerical values of the last oscillation shown in Fig.\  \protect\ref{figVr}(b)
}
\label{figSNtcaseI}
\end{center}
\end{figure}

Simulation results for $S_N(t)$ until $t=1000$ are shown in Fig.\ \ref{figSNtcaseI} for $N=10^3$ and $N=10^6$. Overall good agreement is obtained in the comparison with the theoretical prediction of Eq. (\ref{SNt}), especially when the values $\mu=3/2$ and $A=1.61$  are used. This last finding should  only be taken as the manifestation of the importance of the corrective terms which can lead to such dramatic changes and improvements in $S_N(t)$ after modifying some of the subdominant factors of the survival probability. Of course, additional independent study will be necessary to check the form of $\Gamma_t(r)$ given by Eq.\ (\ref{gasin}) and to find out whether the values for $\mu$ and $A$ considered here are good estimates of the real values \cite{objection}.
Nevertheless, it should be noted that the decrease in $1-\Gamma_t(r)$ when averaging over the whole lattice with respect to its renormalization value, as is implied by the corresponding increment of $\mu$ ($\mu=1/2 \rightarrow \mu=3/2$),  is analogous to the decrease of the  propagator when this same averaging is performed. Given that the two statistical quantities (propagator and survival probability) are closely related, one is inclined  to accept that, at least, the proposed increment in the value of $\mu$  captures the right tendency. The subsequent improvement in the prediction of $S_N(t)$ supports this supposition.

The theoretical expression was not able to give a perfect account of the log-periodic oscillations superimposed on the general trend of $S_N(t)$ shown in Fig.\ \ref{figSNtcaseI}. Notice the self-affinity of the numerical $S_N(t)$ plot (the analytical lines are obviously self-affine): if the segment of horizontal axis of Fig.\ \ref{figSNtcaseI} between $t=0$ and $t=200$ is expanded by a factor $5$ and the corresponding segment of the vertical axis is expanded by a factor $3=5^H$ ($H=d_f/d_w=d_s/2$), we get a figure indistinguishable from Fig. \ref{figSNtcaseI}. 
The origin of these scaling factors is clear: the enlargement of the sides of the Sierpinski gasket by a factor $2$ implies that its fractal volume increases by a factor $3$ and the time that a random walker takes to traverse it increases by $5$ \cite{SHDBA}. 

\begin{figure}
\begin{center}
\parbox{0.4\textwidth}{
\epsfxsize=\hsize \epsfbox{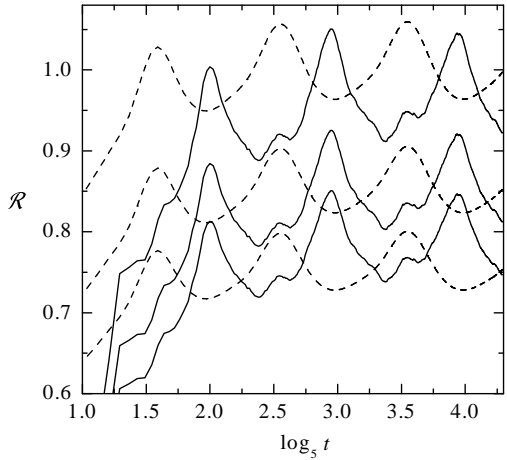}}
\caption{The ratio ${\cal R}$ between theoretical
and numerical values of $S_N(t)$ for the two-dimensional Sierpinski lattice
with $N=10^3$ (dashed line) and $N=10^6$ (solid line).
The asymptotic approximations considered are (from bottom to top) the zeroth-order approximation, the first-order approximation with $\mu=1/2$ and $A=0.61$, and the first-order approximation with $\mu=3/2$ and $A=1.61$. }
\label{figSNtcaseIRatio}
\end{center}
\end{figure}

The structure of $S_N(t)$ is more clearly perceived in Fig.\ \ref{figSNtcaseIRatio} where the quotient between the theoretical prediction and the simulation results is plotted. 
It is remarkable how relatively poor is the performance of the zeroth-order approximation (or main asymptotic term) in predicting the value of $S_N(t)$: it accounts for hardly eighty per cent of $S_N(t)$ for values of $N$ as large as $10^6$. However, the inclusion of the first corrective asymptotic term (especially for some suitable selections of the subdominant parameters $A$ and $\mu$) leads to a noticeable improvement.
 The log-periodic structure is observed both for
$N=10^3$ and $N=10^6$ but in the latter case this structure is richer 
and strikingly similar to that of $V(r)$ as shown in Fig.\ \ref{figVr}(b). We attribute this fact to a better mapping of the lattice structure as more and more random walkers are involved in the exploration. We have plotted the solid line in Fig.\ \ref{figSNtcaseI} with the aim of showing to what extent the oscillatory behavior of $S_N(t)$ as shown in Fig.\ \ref{figSNtcaseIRatio} can be interpreted as a consequence of the oscillatory behavior of $V_0$ shown in Fig.\ \ref{figVr}(b). The line is generated in the same way as the dotted line, i.e., by means of the first-order approximation of Eq.\ (\ref{SNt}) with $A=1.61$ and $\mu=3/2$, but, instead of using the averaged value $V_0=2.93$ (as in the dotted line), we use the actual oscillatory value of $V_0$ taken from the last oscillation (from maximum to maximum) shown in Fig.\ \ref{figVr}(b). The way in which the solid line runs alongside the simulation results supports the view that the log-oscillatory behavior of $S_N(t)$ mainly comes from the log-oscillatory behavior of $V_0$. 

\subsection{Case II}

\begin{figure}
\begin{center}
\parbox{0.4\textwidth}{
\epsfxsize=\hsize \epsfbox{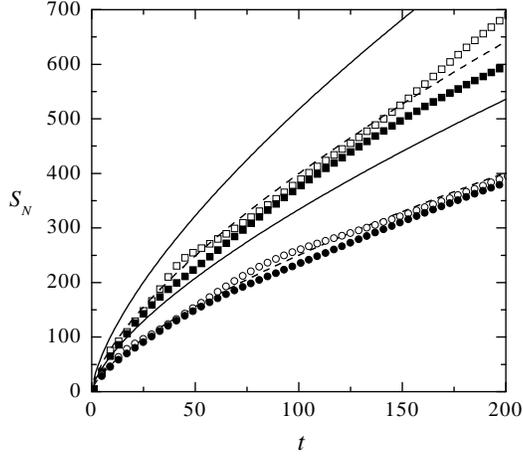}}
\caption{The average number of distinct sites visited in the Sierpinski lattice for $N=2^8$ (filled circles) and $N=2^{13}$ (filled squares) until $t=200$  when the common origin is randomly selected from the shaded area of Fig.\ \protect\ref{figSierpiRed}. The open symbols represent the corresponding values when the origin is the point O. The solid lines are  the zeroth-order approximation and the dashed lines are the first-order approximation with $\mu=3/2$ and $A=1.61$ for, from top to bottom, $N=2^{13}$ and $N=2^8$.
}
\label{figSNtcaseII}
\end{center}
\end{figure}

We also study the effect on $S_N(t)$ of choosing other lattice sites,  besides the point O, as starting sites for the random walkers. To this end we performed simulations where all the $N$ random walkers start on a site randomly selected within the shaded area of Fig.\ \ref{figSierpiRed}  in order to avoid the finite size effects. As expected, the fractal volume $V(r)$ and the average number of distinct sites visited $S_N(t)$ are smooth functions in this case. An estimate of $V_0$ by numerical fitting gives $V_0 \simeq 3.6$. The analysis of the simulation results ($10^4$ runs for $10^3$ randomly selected starting sites) for the mean-square displacement of a single  random walker  is compatible with $2 D \approx 0.8$ when the fit is carried out inside the time interval ($t=50,t=400$). Simulation
results for $S_N(t)$ (five runs for $10^3$ randomly selected starting sites)   until $t=200$ for $N=1024$ are shown in Fig.\ \ref{figSNtcaseII}. They  are compared with the theoretical prediction of the zeroth- and first-order approximations of Eq.\ (\ref{SNt}) for $\mu=3/2$ and $A=1.61$, and with the corresponding simulation results when the origin is at O (case II). Again, we find a relatively poor performance of the zeroth-order approximation, as well as substantial improvement when the fisrt-order approximation is used, although there is still room for further enhancement. It should be noted that the performance of the two asymptotic approximations is completely analogous to that obtained for Euclidean lattices \cite{RC,PREeucl}. 
In these Euclidean media we found that the second-order asymptotic approximation gives rise to a significant improvement in the estimate of $S_N(t)$ even for relatively small values of $N$.  It thus seems natural to conjecture that the same will occur for the Sierpinski gasket, although definitive confirmation of this guess must wait until reliable values for $A$, $\mu$ and $h_1$ are calculated.

\begin{figure}
\begin{center}
\parbox{0.4\textwidth}{
\epsfxsize=\hsize \epsfbox{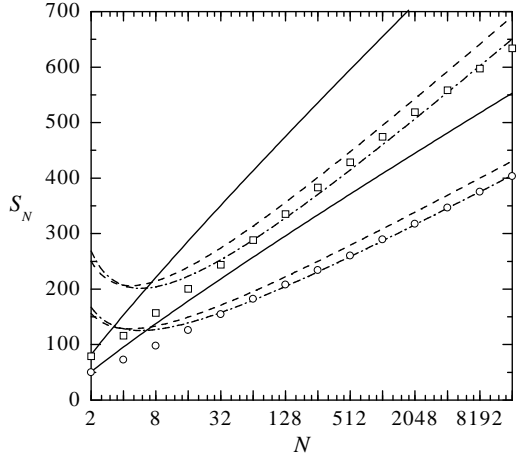}}
\caption{Dependence on $N$ of the fractal territory $S_N(t)$ explored  by $N$ random walkers by time $t=100$ (circles) and $t=200$ (squares) when the common origin is randomly selected from the shaded area of Fig.\ \protect\ref{figSierpiRed}. 
The solid lines are the zeroth-order approximation, and the dashed lines  [dot-dashed lines] correspond to  the first-order approximation with $\mu=3/2$ and $A=1.61$ [$\mu=1.75$ and $A=1.75$] for, from top to bottom, $t=200$ and $t=100$.
}
\label{figSNtversusNcaseII}
\end{center}
\end{figure}

Finally, in Fig.\ \ref{figSNtversusNcaseII} we show the dependence on $N$ of $S_N(t)$ for case II. We have plotted two first-order asymptotic curves: for the first curve we take the usual values  $\mu=3/2$ and $A=1.61$, and the new values  $\mu=1.75$ and $A=1.75$ are used for the second curve.  Again, one sees the great importance of the asymptotic corrective terms as they substantially improve the zeroth-order (main term) asymptotic prediction. We have used the new pair of parameters simply as another example to illustrate the gross effect of the subdominant factors of the survival probability $\Gamma_t(r)$ on the theoretical prediction of $S_N(t)$. The excellent agreement reached  with $\mu=1.75$ and $A=1.75$ should not, however, be considered as an indication that they are the correct  subdominant parameters of the survival probability \cite{objection}.

Notice that the simulation results and theoretical predictions for $S_N(t)$ differ very little from case I to case II despite $V_0$ and $D$ being  clearly different in the two situations. The reason for this coincidence is that the quantity $V_0 (2 D)^{d_f/2}$ that appears in the amplitude of the main asymptotic term of $S_N(t)$ is almost invariant with respect to translations of the origin site: its value was approximately $3.05$ when the origin is placed at O, and  $3.02$ when the origin sites were randomly selected.

\section{SUMMMARY}
\label{sect_4}
We addressed the problem of calculating the average
number $S_N(t)$ of distinct sites visited (or territory covered) by $N$ independent random walkers that diffuse on deterministic fractal lattices. 
The validity of the main result of this paper,  Eq. (\ref{SNt}), is based upon two conditions:
(a) the  volume (number of sites) of the fractal substrate grows as $V(r)=V_0 r^{d_f}$  and 
(b) the asymptotic form of the survival probability, $\Gamma_t(r)$, for large $r$ and small $t$ has essentially the same form as that corresponding
to the Euclidean lattice case. 
The mathematical method used to derive such a result had already been successfully applied to Euclidean lattices \cite{RC,PREeucl} and the fractal case is a fairly straightforward generalization when the previous conditions are fulfilled. In order to check the goodness of the  approximation, we carried out numerical simulations on a standard deterministic substrate (the two-dimensional Sierpinski gasket) obtaining  reasonable agreement with the theoretical results, especially when theoretical first-order asymptotic corrective terms are considered. 
The performance of the theoretical expressions discussed closely resembles that attained for Euclidean lattices.
However, a more definitive check of the theoretical expressions for $S_N(t)$ that include corrective terms is hindered by the uncertainty in the value of the subdominant parameters $A$, $\mu$, $h_1$, \ldots that appear in the survival probability $\Gamma_t(r)$. 
Unfortunately, this function cannot be completely determined by the rigorous renormalization scheme mentioned in Sec.\ \ref{sect_2}, so that the  definitive determination of its subdominant terms by numerical (or 
other analytical procedures) is a problem for future work which will surely be beset by with the technical difficulties associated with the identification of these faint subdominant terms \cite{yusteacedoPRE}.

\acknowledgments
This work has been supported by the DGICYT (Spain) through Grant No. PB97-1501 and
by the Junta de Extremadura-Fondo Social Europeo through Grant No. IPR99C031. 



\begin{references}
\bibitem{objection} A sensible objection to the values $\mu=3/2$ and $A=1.61$ used in Fig.\ \protect\ref{figSNtcaseI} [or $\mu=1.75$ and $A=1.75$ used in Fig.\ \protect\ref{figSNtversusNcaseII}] is that they lead to a first-order approximation which is so good that almost no room is left for further improvements coming from higher-order asymptotic terms.
\end{references}
\end{document}